\documentclass[prl,twocolumn,showpacs,amsmath,amssymb,superscriptaddress]{revtex4}
\usepackage{pifont,amssymb,epsf,graphicx}
\begin{document}

\title{Tunneling interstitial impurity in iron-chalcogenide based superconductors}

\author{Huaixiang Huang}
\affiliation{Shanghai Key Laboratory of High Temperature Superconductors, Department of Physics, Shanghai University, Shanghai 200444, China}
\affiliation{Texas Center for Superconductivity and Department of
Physics, University of Houston, Houston, Texas 77204, USA}

\author{Degang Zhang}
\affiliation{Texas Center for Superconductivity and Department of
Physics, University of Houston, Houston, Texas 77204, USA}
\affiliation{College of Physics and Electronic Engineering, Sichuan Normal University, Chengdu 610101, China}
\author{Yi Gao}
\affiliation{Department of Physics and institute of Theoretical Physics, Nanjing Normal University, Nanjing, Jiangsu 210023, China}

\author{Wei Ren}
\affiliation{International Centre for Quantum and Molecular Structures, Materials Genome Institute, Shanghai University, Shanghai 200444, China}
\affiliation{Shanghai Key Laboratory of High Temperature Superconductors, Department of Physics, Shanghai University, Shanghai 200444, China}

\author{C. S. Ting}
\affiliation{Texas Center for Superconductivity and Department
of Physics, University of Houston, Houston, Texas 77204, USA}

\date{\today}

\begin{abstract}
A pronounced local in-gap zero-energy bound state (ZBS) has been observed by recent scanning tunneling microscopy (STM) experiments on the interstitial Fe impurity (IFI) and its nearest-neighboring (nn) sites in $\mathrm{FeTe_{0.5}Se_{0.5}}$ superconducting (SC) compound. By introducing a new impurity mechanism, the so-called tunneling impurity, and based on the Bogoliubove-de Gennes (BDG) equations we investigated the low-lying energy states of the IFI and the underlying Fe-plane.
We found the peak of ZBS does not shift or split in a magnetic field as long as the tunneling parameter between IFI and the Fe-plane is sufficiently small and the Fe-plane is deep in the SC state. Our results are in good agreement with the experiments.
We also predicted that modulation of spin density wave (SDW), or charge density wave (CDW) will suppress the intensity of the ZBS.
\end{abstract}
\maketitle

Since the discovery of iron-based superconductor,~\cite{kam} new compounds continue to be found. The 11 type iron chalcogenides has attracted much attention due to the simplicity of its crystal structure. Angle-resolved photoemission spectroscopy, STM, and transport experiments~\cite{11stm,feseaps,multgap3,transp} have been performed to investigate the electronic structure and SC gap. By substituting Se for Te in FeTe compound, superconductivity appears and presents a variety of phenomena for different mixing ratios of Te and Se~\cite{w11}.
The as-grown $\mathrm{Fe_{1+y}Te_xSe_{1-x}}$ single crystals contain a large amount of excess Fe that are randomly situated at the interstitial sites in the crystal. Recent STM experiments investigated the electronic state near the interstitial Fe in $\mathrm{FeSe_{0.5}Te_{0.5}}$ superconductors.~\cite{yjx}
A robust ZBS is observed in the tunneling spectrum taken at the center of the IFI and its nn sites with the intensity decaying quickly.
Interestingly, the peak does not split or shift in a magnetic field which is drastically inconsistent with the magnetic or non-magnetic impurity effects in either d-wave or s-wave superconductors.~\cite{2,3,4}

The interplay of the IFI with the in-plane Fe and the opposite Se (Te) are complex. The valence of excess iron does not equal to that of the in-plane iron, the effects of the IFI to the physical property are sensitive to the stoichiometry.~\cite{feseaps,mag3,mag1,mag2,bicoll}.
Although the magnetic moment of the access Fe has finite value, for SC state $\mathrm{Fe_{1+y}Te_{0.5}Se_{0.5}}$,
we assume that the IFI is coupled to the underlying Fe-plane by a hopping term. Taking the IFI as a tunneling impurity~\cite{zhangdg} without scattering potential, and based on the BDG equations in the presence as well as in the absence of the magnetic field, we systematically investigated the low-lying energy states of the IFI and the in-plane Fe. We found the IFI-induced in-gap ZBS is a common feature for iron-based superconductors as long as the hopping term between the IFI and the underlying Fe-plane is sufficiently small and Fe-plane is deep in the SC state. The ZBS is isotropic and localized,
the intensity of the induced ZBS decays to zero at a distance of three or four lattice constants away from the IFI.

In the optimally doped regime, and when the external magnetic field is not very large, beside the appearance of the in-gap resonance peaks induced by the magnetic field~\cite{gao2,zou3}, the IFI-induced no-splitting no-shifting ZBS shows up at its nn sites no matter where we put the IFI. This phenomenon is contrary to the Zeeman effect of single electron since the IFI-induced ZBS results from Cooper pairs. Intensity of the ZBS at nn sites has the maximum value as the IFI is located above the vortex core center.
Considerably large magnetic field may induce charge oscillation, and breaks the process of reforming Cooper pairs when electron tunnels back to the Fe-plane from the IFI. Therefore larger magnetic field will suppress the ZBS at the nn sites of the IFI.  We believe that the tunneling impurity is a new scenario to understand the experimental finding of IFI effect.

\begin{figure}
\includegraphics[width=5cm]{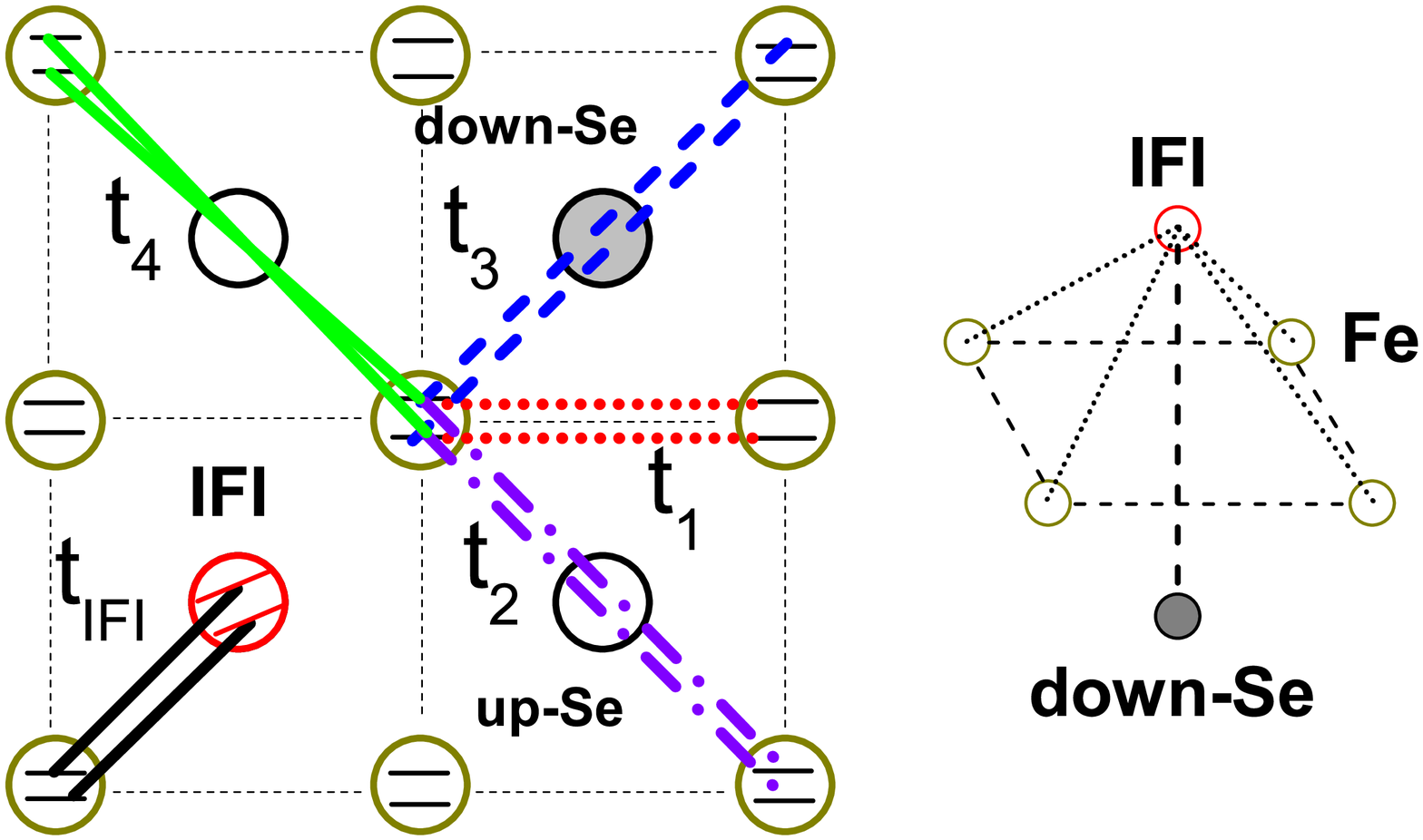}
\caption{(color online)Schematic plot of the tight-binding model. The right picture shows the relative position of IFI to the Fe-plane. }\label{f1}
\end{figure}
Since the electronic structure and Fermi-surface topology of the 11 system is very similar to those of the other iron-based superconductors,~\cite{feseaps} we use a minimal two-orbital four-band tight binding model~\cite{zhang} to investigate $\mathrm{Fe_{1+y}Te_{0.5}Se_{0.5}}$. Different from the co-planar $\mathrm{CuO_2}$, superconducting Fe-plane is sandwiched between two adjacent $\mathrm{Se(Te)}$ layers, the next-nearest-neighbor (nnn) hoppings are mediated by up and down Se (Te) and are not equal to each other. This asymmetry should be correct when one investigates the surface properties as in STM experiments, because the bonds between up Te (Se) ions and Fe ions are broken after cleavage.
Theoretical results~\cite{gao2,zou3,zho,gao1,huang,huang1,huang2} based on this model are qualitatively consistent with
experiments.~\cite{arpes,nematic,jyong}
The projected up and down Se (Te) atoms stay at the centers of the Fe plaquette alternatively, and the IFI is located at the opposite side of the down Se (Te) above the Fe-plane[see Fig.\ref{f1}].
The total Hamiltonian is expressed as
$
H=H_{0}+H_{\mathrm{IFI}}$, $H_{0}$ is the mean-field phenomenological model Hamiltonian without excess Fe, reads
\begin{eqnarray}\label{h0}
H_{0}&=&-\sum_{{\bf i}\mu {\bf j}\nu\sigma}(t_{{\bf i}\mu
{\bf j}\nu}c^\dagger_{{\bf i}\mu\sigma}c_{{\bf j}\nu\sigma}+h.c.)-\tilde{\mu}\sum_{{\bf i}\mu\sigma}c^{\dagger}_{{\bf i}\mu\sigma}c_{{\bf i}\mu\sigma}
\nonumber\\&+&U^{\prime}\!\!
\sum_{{\bf i},\nu,\sigma}\!\!\langle
n_{\bf i}\rangle n_{{\bf i}\nu\sigma}+\!\!\!
\sum_{{\bf i}\mu {\bf j}\nu\sigma}\!\!\!(\Delta_{{\bf i}\mu
{\bf j}\nu}c^\dagger_{{\bf i}\mu\sigma}c^{\dagger}_{{\bf j}\nu\bar{\sigma}}+h.c.)\;,
\end{eqnarray}
where $n_{{\bf i}\mu\sigma}$ is the electron concentration at site ${\bf
i}$, orbital $\mu$ for spin $\sigma$, $\tilde{\mu}$ is the chemical potential, determined by the average electron filling on Fe site $\langle n\rangle_i=2+x$, $x$ denotes electron doping concentration. $H_{0}$ can be separated to three parts. The first line is tight binding part, with the hopping integral  $t_{1-4}=1, 0.4, -2, 0.04$~\cite{zhang} as depicted in Fig.\ref{f1}.
The first term in the second line is the mean field expression of Coulomb and Hund's interaction $H_{0,int}$ in the SC state without SDW or CDW. It will be more complicated when SDW or CDW exists
\begin{eqnarray}\label{h0int}
H_{0,int}&=&\sum_{{\bf i},\nu^{\prime}\neq\nu,\sigma}U\langle
         n_{{\bf i}\nu\bar{\sigma}}\rangle n_{{\bf
         i}\nu\sigma}+ (U-3J_H)\langle
          n_{{\bf i}\nu^{\prime}\sigma}\rangle n_{{\bf i}\nu\sigma}\nonumber\\
        &+&(U-2J_H)\langle n_{{\bf
i}\nu^{\prime}\bar{\sigma}}\rangle n_{{\bf i}\nu{\sigma}}.
\end{eqnarray}
Here we take $U=3.4$, $J_H=1.3$ and $U^{\prime}=\frac{3U-5J_H}{4}$.~\cite{zho,gao1,huang,huang1,huang2} The last term in the second line is the phenomenological superconducting pairing part. Although there are controversies about the symmetry of the SC pairing, $s_{\pm}$-wave~\cite{multgap1,multgap2,multgap3} is suggested for $\mathrm{Fe_{1+y}Te_{1-x}Se_x}$. In real space it is intraorbital pairing $\Delta_{{\bf i}\mu {\bf j}\nu}= \frac{V}{2}\langle
c_{\bf{i}\mu\uparrow}c_{\bf{j}\mu\downarrow}-c_{\bf{i}\mu\downarrow}c_{\bf{j}\mu\uparrow}\rangle$, where $j$ is the nnn of $i$ site. The self-consistent mean field are $\Delta_{{\bf i}\mu {\bf j}\mu}$ and $\langle n_{{\bf i}\mu}\rangle$.

The role of the IFI is to provide two tunneling channels corresponding to the two orbitals of Fe. Electron tunneling onto IFI belongs to a Cooper pair since the Fe-plane is in the SC state, when it tunnels back to the Fe-plane a new Cooper pair reforms. Because the effective magnetic moment of the cooper pair is zero, there is no magnetic interaction between the IFI and the Fe-plane. The effective coupling of the IFI to the Fe-plane reads
\begin{eqnarray}\label{h2}
H_{\mathrm{IFI}}=-t_{\mathrm{IFI}}\sum_{\langle \bf{i}\mu\sigma\rangle} \tilde{c}^\dagger_{\mu\sigma}c_{{\bf i}\mu\sigma}\, ,
\end{eqnarray}
where $\tilde{c}^{\dag}$ is the creation operator of the IFI, $\langle \rangle$ means the summation up to its four nn sites in the Fe-plane. Tunneling magnitude $t_{\mathrm{IFI}}$ is a tuning parameter in our calculations.

Magnetic field contains a dynamical term for electrons, in the mixed state, the effect of the magnetic field is included through the Peierls phase factor. For a perpendicular magnetic field, the hopping integral in Eq.(\ref{h0}) and (\ref{h2}) should be changed to $t^{\prime}$ which can be expressed as
$t^{\prime}_{\bf{i}\mu \bf{j}\nu}=t_{\bf {i}\mu \bf {j}\nu} \exp{[\texttt{i}\varphi_{ij}]}$, where
$\varphi_{ij}=\frac{\pi}{\Phi_0}\int^i_j \textbf{A}(\textbf{r})\cdot d\bf{r}$, and $\Phi_0=hc/2e$ is the superconducting flux quantum. We assume the applied magnetic field B to be uniform and the vector potential is $\textbf{A}=(-\mathrm{B}y,0,0)$ in the Landau gauge. Periodic boundary condition should be consistent with the requirement that the total phase factor along each small plaquette given by $\sum_{\square}\varphi_{ij}=\frac{\pi B a^2}{\Phi_0}$.

The presence of Peierls phase makes the usual translation operator change to magnetic translation operator.
To ensure them commutable with each other and the Hamiltonian, each magnetic unit cell has to contain  $2\Phi_0$ flux~\cite{zjxm,cym}. In this case the supercell technique can be used to calculate the local density of states (LDOS) in the presence of B.
We self-consistently solve the BDG equations in real space, linear dimension of a unit cell is $N_x\times N_y=32\times 32$ for $\mathrm{B}=0$, while for finite B we adopt $N_x=2N_y$. The LDOS is calculated according to
\begin{equation}
\rho_{\bf i}(\omega)=\sum_{n\mu {\bf k}}[|u^{n}_{{\bf i}\mu\sigma{\bf k}}|^{2}\delta(E_{n, {\bf k}}-\omega)+
|v^{n}_{{\bf i}\mu\bar{\sigma}{\bf k}}|^{2}\delta(E_{n, {\bf k}}+\omega)],\nonumber
\end{equation}
where the delta function $\delta(x)$ is
$\Gamma/\pi(x^2+\Gamma^2)$, with the quasiparticle damping
$\Gamma=0.005$. $u^n_{\bf i \mu {\bf k} }$ and $v^n_{\bf i \mu {\bf k}}$ are the eigenstate component of the energy $E_{n,{\bf k}}$ at site $\bf i$ orbital $\mu$ for wave vector ${\bf k}$, and they correspond to the particle-like component and hole-like component respectively. The number of unit cell is taken as $M_x\times M_y=20\times 20$ for $\mathrm{B}=0$ and $10\times 20$ when external B is applied, with $\bf {k}_{\alpha}=-\frac{\pi}{N_{\alpha}}+\frac{j2\pi}{N_{\alpha}M_{\alpha}}$, $\alpha=x,y$.
Throughout the paper, temperature is zero K, the energy and length are measured
in units of $t_{1}$ and the nearest Fe-Fe distance $a$, respectively.

\begin{figure}
\includegraphics[width=8cm]{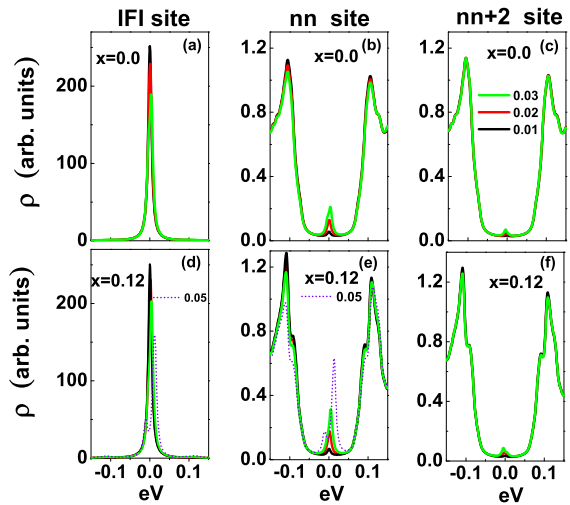}
\caption{(color online) LDOS at IFI and its nn sites as well as at the nearby $nn+2$ site for different $t_{\mathrm{IFI}}=0.01t_1,0.02t_1,0.03t_1$. Panel (a)-(c) are for $x=0.0$ while panel (d)-(f) for $x=0.12$. In (d) and (e) the dotted volet line are for $t_{\mathrm{IFI}}=0.05$.} \label{f3}
\end{figure}
Although the experiments was performed on half-filled SC compounds~\cite{yjx}, we found ZBS is a common feature for SC state as long as $t_{\mathrm{IFI}}$ is small. Fig.\ref{f3} shows LDOS $\rho$ for different $t_{\mathrm{IFI}}$ with $x=0.0$ and $x=0.12$ for the Hamiltonian of Eq.\ref{h0}. For $x=0.0$ and $t_{\mathrm{IFI}}$ less than $0.04$, LDOS at the IFI has a single sharp ZBS peak which can be seen in Fig.\ref{f3}(a), and the peak height is decreased with the increasing of $t_{\mathrm{IFI}}$ .

At its four nn sites $\rho$ is isotropic. Compared to the pure sample, a pronounced in-gap ZBS appears and the height of the peak is enhanced with $t_{\mathrm{IFI}}$ increasing from $0.01$ to $0.03$ as seen in Fig.\ref{f3}(b). At the sites two lattice constants away from the nn sites, the intensity of the in-gap ZBS decays almost to zero as shown in Fig.\ref{f3}(c).
For different doping cases, we expect the ZBS still exist whenever Fe-plane is in the SC state. Fig.\ref{f3}(d)-(f) plot $\rho$ at the IFI and its vicinity for $x=0.12$. The curves are very similar to those of the $x=0.0$ cases
with the corresponding peak of the ZBS a little higher.

 Small and large $t_{\mathrm{IFI}}$ give significantly distinct results. For relatively larger $t_{\mathrm{IFI}}$, the results deviate from the experimental observation.
When $x=0.12$, $t_{\mathrm{IFI}}=0.05$, $\rho$ at the IFI splits into two asymmetrical peaks shown by the violet dotted line in Fig.\ref{f3}(d). Due to the proximity effect the LDOS at the nn sites of the IFI also shows double peaks with heavy weight on the positive energy and can be seen in Fig.\ref{f3}(e). Therefore, sufficiently small $t_{\mathrm{IFI}}$ is an important condition for the appearance of ZBS. As temperature increases, the phonon mediated  layer tunneling $t_{\mathrm{IFI}}$ is also increased, thus ZBS will not appear for higher temperature cases.

\begin{figure}
\includegraphics[width=8cm]{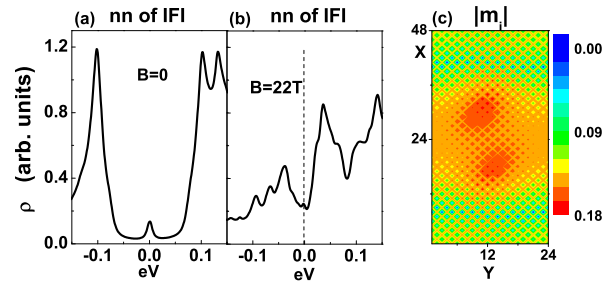}
\caption{(color online) For $t_{\mathrm{IFI}}=0.02$, $x=0.06$, (a)LDOS at nn sites without B. (b)LDOS at nn site for $\mathrm{B\thickapprox22T}$ with IFI projected at vortex core center $(12.5,12.5)$. (c)Imagine of modulated $|m_i|$ for $\mathrm{B\thickapprox22T}$ with IFI above vortex core center.} \label{f7}
\end{figure}
We have addressed the cases when Fe-plane does not have magnetic order defined as $m_i=(n_{i\uparrow}-n_{i\downarrow})/2$. In fact, as magnetic order coexists with the SC order which will happen in underdoped cases, the LDOS at the IFI and its nn sites will still have the in-gap ZBS. In the following we will discuss the more complicated cases with $H_{0,int}$ taking the form of Eq.\ref{h0int}.  The formation of the ZBS at the nn sites is similar to the Andreev reflection and we expect the homogeneous magnetic order itself does not suppress ZBS as long as the sample is in the SC state. We can see from Fig.\ref{f7}(a) that the ZBS still appears at nn sites of the IFI for underdoped case $x=0.06$ with $t_{\mathrm{IFI}}=0.02$. While an external B could drive SDW modulation [see Fig.\ref{f7}(c)] and CDW in this case, the competition between the oscillated SDW and SC order ruins the proximity effect. Therefore as SDW exists, an external magnetic field will suppress the ZBS as well as other in-gap states which is depicted in Fig.\ref{f7}(b). In Fig.\ref{f7}(b), the IFI is located above the vortex core center with the ZBS almost vanished; as the above IFI moves away from the center, the ZBS does not appear.
\begin{figure}
\includegraphics[width=8.5cm]{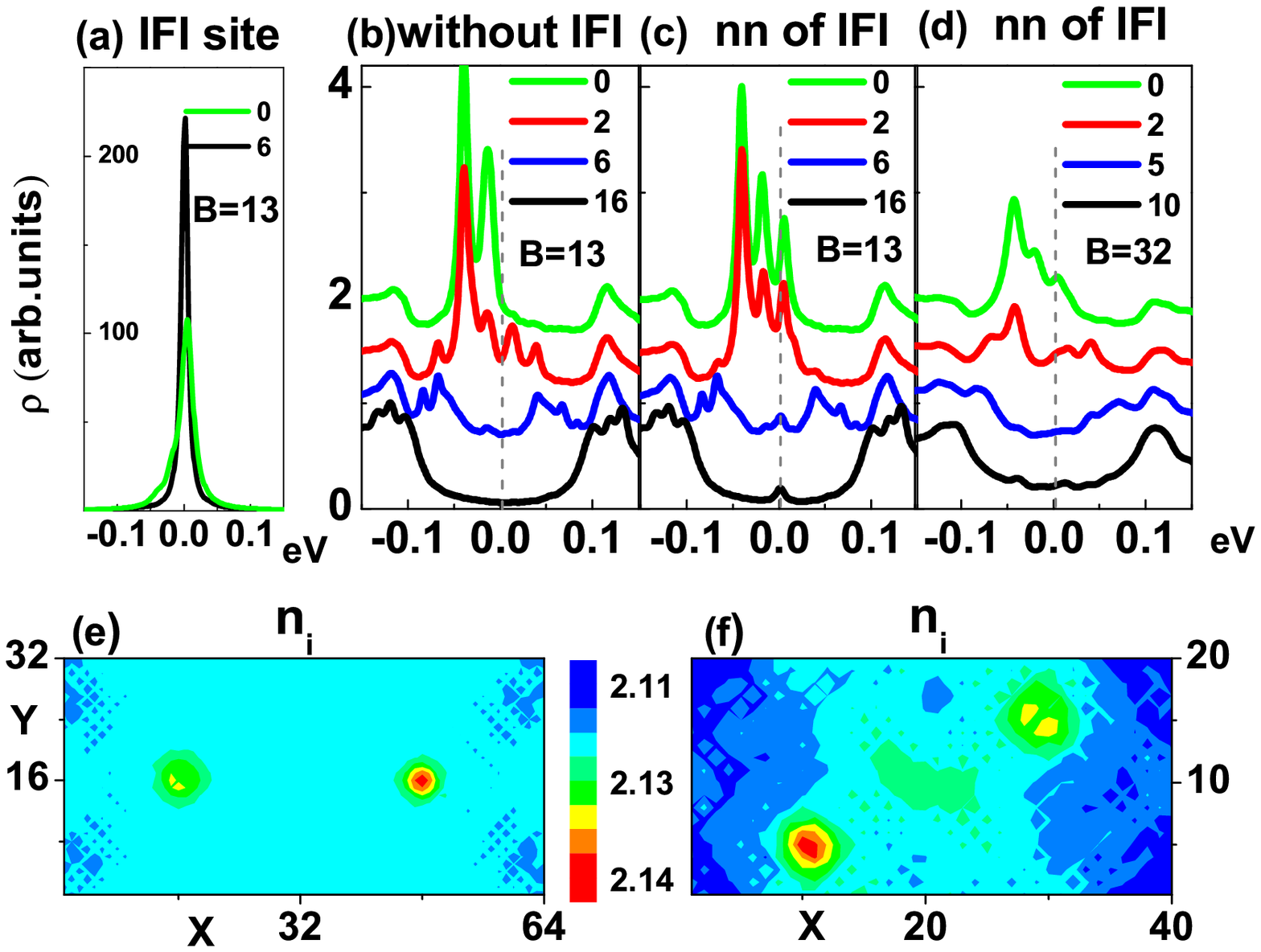}
\caption{(color online)(a)LDOS at different IFI for $\mathrm{B\thickapprox13T}$.(b)Without IFI, LDOS at vortex core center and at sites away from vortex for $\mathrm{B\thickapprox13T}$.(c)With the IFI located at different place, LDOS at the corresponding nn site for $\mathrm{B\thickapprox13T}$. (d)Similar to (c) but for $\mathrm{B\thickapprox32T}$. From up to down the curves correspond to the projected IFI at vortex core center, and L lattice constants away from the center along horizontal axis. The gray dashed line indicate the position of zero energy. (e)Image of $n_i$ for $\mathrm{B\thickapprox13T}$ with the projected IFI at $(16.5,16.5)$. (f)Image of $n_i$ for $\mathrm{B\thickapprox32T}$ with the projected IFI at $(29.5,15.5)$. The curves of (b)(c)and (d) are displaced vertically by 0.5 unit for clarify. } \label{f6}
\end{figure}

In the optimally doped regime, we investigated perturbation of IFI on the mixed state.
$\rho$ is plotted in Fig.\ref{f6} at IFI as well as at its nn sites for different B with fixed $t_{\mathrm{IFI}}=0.02$, $x=0.12$. A unit cell of linear dimension $64\times 32$ corresponds to a relatively small magnetic field $\mathrm{B\thickapprox13T}$ and $40\times 20$ corresponds to relatively large field $\mathrm{B\thickapprox32T}$.
In all cases of small $t_{\mathrm{IFI}}$, $\rho$ on the IFI has a sharp ZBS regardless of the strength of B and the position of the IFI. The height of the peak becomes lower as the IFI moves close to the vortex core center which can be seen in Fig.\ref{f6}(a).

Now we look at the relatively small magnetic field $\mathrm{B\thickapprox13T}$. In the absence of the IFI, two in-gap resonance peaks induced by B are located at negative energies in the vortex core. At sites away from the vortex, resonance peaks are suppressed and move to the gap edge, finally evolve into its bulk feature [see Fig.\ref{f6}(b)]. When introducing IFI into the system, the above characteristics do not change except for the additional appearance of ZBS induced by the IFI which is depicted in Fig.\ref{f6}(c). We can see that the in-gap ZBS shows up at nn sites no matter where we put the IFI. Compared with the cases without B [Fig.\ref{f3}], the height of the ZBS at nn sites of the IFI is obviously enhanced, especially when the IFI is located above the vortex core center and has its maximum value. As the IFI moves away from the vortex center, the peak on nn sites is decreased. When the projected point of IFI on Fe-plane is $16$ lattice constants away from vortex core center along $x$-axis, the height of the ZBS of nn sites evolves back to the value of $\mathrm{B}=0$ case. We also notice that a lower height of ZBS on the IFI [see Fig.\ref{f6}(a)] corresponds to a higher peak of ZBS on its nn sites [see Fig.\ref{f6}(c)]. The more of quasiparticle tunneling onto nn site, the less part is remained on the IFI. Results of $\mathrm{B\thickapprox22T}$ ($48\times24$ unit cell) are very similar to those of $\mathrm{B\thickapprox13T}$ and we do not show them here.

Without the IFI, $n_i$ is enhanced and the pairing parameter $\Delta_i=\sum_j \Delta_{ij}/4$ is almost vanished at the vortex core center, with $\Delta_i$ and $n_i$ being symmetric with respect to the two vortex cores.
Introducing of the IFI breaks the symmetry. For smaller $\mathrm{B\thickapprox13T}$ and projected IFI at vortex core center $(16.5,16.5)$, although $n_i$ on the two vortex core are different, it increases to bulk value at the scale of coherence length. Out of vortex core, $n_i$ is almost homogeneous which can be seen in the Fig.\ref{f6}(e). While larger B leads to oscillation of $n_i$, Fig.\ref{f6}(f) displays the inhomogeneous $n_i$ for $\mathrm{B\thickapprox32T}$ with IFI located above vortex core center $(29.5,15.5)$. The difference of $n_i$ on the two vortex cores is $0.01$, much larger than $\mathrm{B}=0$ cases in which $\Delta n_i$ induced by IFI is at the order of $10^{-3}$. Obviously, magnetic field significantly enhanced the impact of IFI to the Fe-plane. The oscillated CDW ruins the Cooper pair tunneling and reforming, hence
suppresses the ZBS on nn sites. For large $\mathrm{B\thickapprox32T}$, Fig.\ref{f6}(d) shows that when IFI locates above vertex core center, ZBS appears at its nn sites with the peak much lower than that of small B case. As the above IFI moves away from vortex core center, it still have impact on nn sites, however the induced state is not ZBS and almost invisible in Fig.\ref{f6}(d).

In summary, motivated by the recent STM experiments~\cite{yjx} on iron-based superconductor $\mathrm{FeTe_{0.5}Se_{0.5}}$, we investigated the impact of IFI on the $s_{\pm}$-wave iron-based superconductors. Taking IFI as a tunneling impurity,~\cite{zhangdg} we calculated LDOS on the IFI as well as on Fe-plane in the absence and in the presence of magnetic field.
In all cases with and without external magnetic field, LDOS at the IFI has a single sharp ZBS as long as hopping parameter $t_{\mathrm{IFI}}$ is sufficiently small and Fe-plane is in SC state.
The ZBS always appears on its nn sites due to the proximity effect in the absence of B.
When a magnetic field is applied, our results are in good agreement with the experiments. When Fe-plane is deep in SC state, the IFI-induced ZBS shows up at nn sites no matter where we put the IFI. It does not split or shift and the intensity is enhanced. The effect of the IFI is isotropic and localized in all cases.
However when Fe-plane has oscillated SDW or CDW, the intensity of ZBS on nn sites will be suppressed.
In underdoped cases, a magnetic field can lead to modulation of SDW.
And when the applied magnetic field is considerably large, it may induce CDW in the IFI system.

Tunneling impurity only provide tunneling channels in the system. Due to the superconducting state of Fe-plane,
the effective tunneling is Cooper pair tunneling with zero magnetic moment thus no Zeeman splitting. Although experiments and density functional theory study verified the access Fe is spin polarized\cite{mag1}, we do not consider exchange interaction between excess Fe and superconducting Fe-plane. The origin of IFI-induced ZBS on nn sites is related to Andreev reflection.
As far as we know, a convinced explanation for the STM experiments is still lacking. Although it may associated with non-trivial topological or other factors, our simple tunneling IFI gives a natural explanation for the experiments and provides a new interpretation for interstitial impurity.

\begin{acknowledgements}
We thank Jiaxin Yin for helpful discussions.
This work was supported by the Texas Center for Superconductivity at the University of Houston and the Robert A. Welch Foundation under the Grant No. E-1146,
National Key Basic Research Program of China (Grant No. 2015CB921600), QiMingXing Project (No. 14QA1402000) of Shanghai Municipal Science and Technology Commission, Eastern Scholar Program and Shuguang Program (No. 12SG34) from Shanghai Municipal Education Commission,
 NSF of Shanghai(Grant No. 13ZR1415400), Shanghai Key Lab for Astrophysics (Grant No. SKLA1303), NSFC (Grant Nos. 11204138 and 11274222) and
NSF of Jiangsu Province of China (Grant No. BK2012450),
the Sichuan Normal University and the "thousand talent program" of Sichuan Province, China.
\end{acknowledgements}

\end{document}